\documentclass[%
superscriptaddress,
aps,
reprint,%
amssymb, amsmath,%
groupedaddress,%
prl,
longbibliography,
]{revtex4-1}
\usepackage{graphicx}%
\usepackage{docs}%
\usepackage{bm}%

\usepackage[colorlinks=true,linkcolor=magenta, citecolor=blue]{hyperref}%

\newcommand{\aSi}{\mbox{$\alpha$-Si}}

\makeatletter
\@fpsep\textheight
\makeatother

\begin{document}

\title{Wide bandwidth and high resolution planar filter array based on DBR-metasurface-DBR structures}




\author{Yu Horie}
\affiliation{T. J. Watson Laboratory of Applied Physics, California Institute of Technology, 1200 E. California Blvd., Pasadena, CA 91125, USA}
\author{Amir Arbabi}
\affiliation{T. J. Watson Laboratory of Applied Physics, California Institute of Technology, 1200 E. California Blvd., Pasadena, CA 91125, USA}
\author{Ehsan Arbabi}
\affiliation{T. J. Watson Laboratory of Applied Physics, California Institute of Technology, 1200 E. California Blvd., Pasadena, CA 91125, USA}

\author{Seyedeh Mahsa Kamali}
\affiliation{T. J. Watson Laboratory of Applied Physics, California Institute of Technology, 1200 E. California Blvd., Pasadena, CA 91125, USA}

\author{Andrei Faraon}
\email{Corresponding author: A.F: faraon@caltech.edu}
\affiliation{T. J. Watson Laboratory of Applied Physics, California Institute of Technology, 1200 E. California Blvd., Pasadena, CA 91125, USA}

\begin{abstract}
We propose and experimentally demonstrate a planar array of optical bandpass filters composed of low loss dielectric metasurface layers sandwiched between two distributed Bragg reflectors (DBRs). The two DBRs form a Fabry-P\'{e}rot resonator whose center wavelength is controlled by the design of the transmissive metasurface layer which functions as a phase shifting element. We demonstrate an array of bandpass filters with spatially varying center wavelengths covering a wide range of operation wavelengths of
$250\,\mathrm{nm}$ around
$\lambda=1550\,\mathrm{nm}$
($\Delta\lambda/\lambda=16\%$). The center wavelengths of each filter are independently controlled only by changing the in-plane geometry of the sandwiched metasurfaces, and the experimentally measured quality factors are larger than 700.
The demonstrated filter array can be directly integrated on top of photodetector arrays to realize on-chip high-resolution spectrometers with free-space coupling.
\end{abstract}

\maketitle


\section{Introduction} 
\label{sec:introduction}

Spectroscopy is an essential tool in bio-chemical sensing applications, material characterization, and multiple areas of scientific research. Modern spectrometers based on diffraction gratings are widely used because they can achieve a high resolving power and high sensitivity.
For multiple applications, including those related to sensors located on handheld devices and low cost portable point-of-care diagnostics \cite{Minas:2005cz}, there is a continuous interest in miniaturizing spectrometers.
However, conventional high resolution diffraction grating based spectrometers are inevitably bulky as the resolution of the spectrometer scales inversely with optical path length, and thus are not suitable for miniaturization.
For this purpose, several integrated optics approaches have been explored \cite{Momeni:2009ef,Bacon:2004cs,Wolffenbuttel:2005bs}, such as on-chip frequency filtering based on micro-resonators \cite{Xia:2011wf}, integrated diffraction gratings \cite{Kyotoku:2010tk}, and arrayed waveguide gratings \cite{Cheben:2007vn}.
However, in many applications the optical signals of interest are freely propagating, and the low coupling efficiency from free-space to on-chip waveguides limits the sensitivity of this type of spectrometers.
An attractive design for a free-space coupled spectrometer is to use an array of bandpass optical filters in conjunction with a photodetector array \cite{Wang:2007ty,Correia:2000je}.
One can obtain the spectral information by measuring intensities of the filtered light within a specific range of wavelengths at each detector, and more importantly the resolving power of the spectrometer is only limited by the resolution of the filters.
The most common way to design a high-resolution optical filter is to form a Fabry-P\'{e}rot (FP) resonator using a pair of broadband high reflectivity mirrors \cite{Correia:2000je}. The FP cavity length can be varied in a discrete form through multiple etching steps, or in a continuous form by using an angled surface.
The latter creates optical filters with spatially varying center wavelength, named wedge filters, that are manufacturable by linearly varying the cavity thicknesses of the FP resonator \cite{Emadi:2012tv}, and are commercially available \cite{OBrien:2012fn}.
However, the angle of the wedge limits the quality factor of the FP cavities and in turn the resolution of the filters due to the non-normal reflection on the angled surface.
Gray-scale lithography enables a spatially varying cavity thicknesses in a more controlled manner \cite{Xiao:2012fp}, but the technology is expensive and not readily available.

\begin{figure}[p!]
  \centering
  \includegraphics{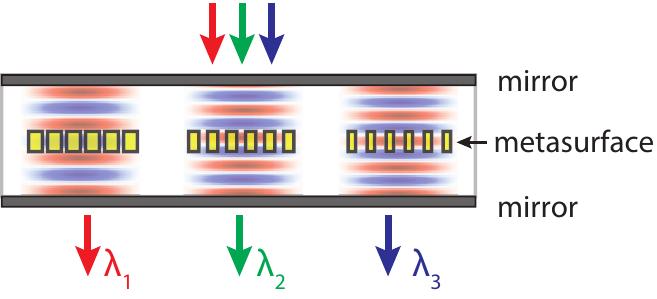}
  \caption{Schematic of the proposed bandpass filter array composed of vertical DBR-based micro-cavities, in which transmissive dielectric metasurface layers are inserted as phase shifting layers to tune their resonance wavelengths over a broad bandwidth.}
  \label{1_concept}
\end{figure}

In this article, we propose and experimentally demonstrate a new method to effectively vary the central wavelengths of a FP filter set by inserting a transmissive dielectric metasurface as a phase shifting element between two high reflectivity mirrors, enabling independent and precise control of the filter's passbands.
Metasurfaces are two dimensional arrays of subwavelength structures capable of controlling the phase, amplitude, and polarization of light \cite{Yu:2014hq,Jahani:2016fq,Kildishev:2013hq}.
One particularly interesting class of metasurfaces are based on high index nano-posts surrounded by a low index medium, which allow both high transmission as well as phase control capability by designing the geometry of the nano-posts.
So far, various diffractive optical elements such as high performance flat lenses \cite{Vo:2014jl,Lin:2014ht,Arbabi:2015iq} or birefringent optical elements \cite{Arbabi:2015foa} have been demonstrated.
Unlike plasmonic metasurfaces which inevitably suffer from optical loss \cite{Khurgin:2015fq}, the loss-less nature of dielectric metasurfaces is suitable for resonant applications.
As schematically shown in Fig.~\ref{1_concept}, the dielectric metasurface layers are incorporated in vertical FP resonators with relatively high quality factors.
By incorporating transmissive metasurfaces with different geometries
into the cavity of a set of FP filters, the round-trip phase inside the cavity is drastically modified.
Thus, the resonance wavelength (\textit{i.e.} the filter passband) can be tuned without changing the physical distance between the two mirrors.
Similar concepts for implementation of arrays of FP filters have been previously studied. Walls et al. have demonstrated FP filter arrays using metallic mirrors and effective index medium created by subwavelength patterning  \cite{Chen:2012jd}.
Filter arrays composed of dielectric mirrors and 1D subwavelength gratings have also been proposed \cite{Kaushik:1995kha}, but, to the best our knowledge, have not been experimentally demonstrated. Furthermore, compared with 1D subwavelength gratings, the dielectric metasurfaces provide more control over the phase shifts and are polarization insensitive ~\cite{Vo:2014jl,Arbabi:2015iq,Arbabi:2014tp}.



\section{Design} 
\label{sec:design}

\begin{figure}[p!]
  \centering
  \includegraphics{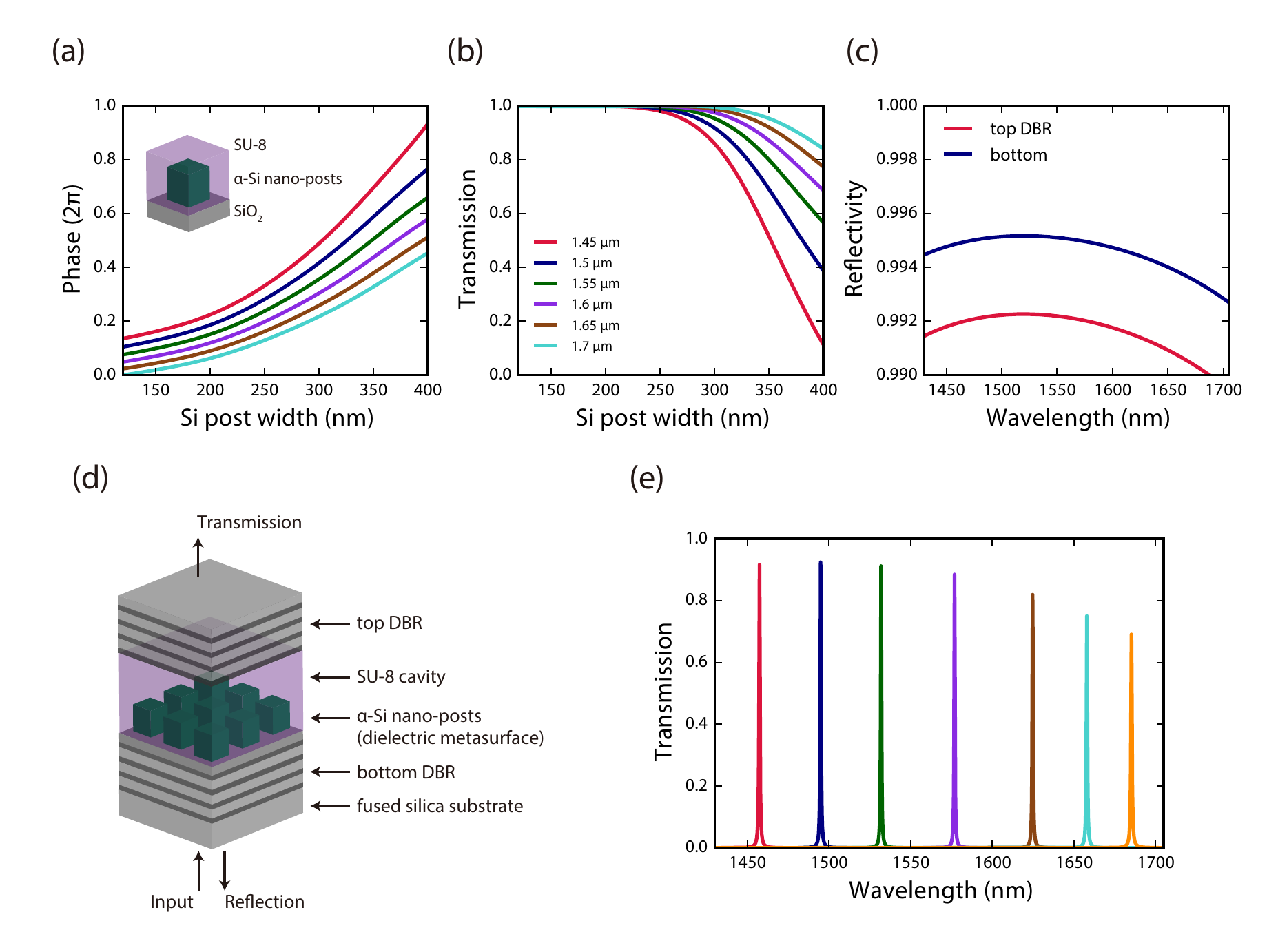}
  \caption{(a) Transmission round-trip phase, and (b) transmission intensity induced by \aSi{} \mbox{nano-posts} as a function of post width for different wavelengths.
  The inset figure in (a) represents the refractive index profile of the dielectric metasurface considered.
  (c) The simulated reflection spectrum of DBRs.
  (d) Schematic illustration of the proposed filters. The filters are composed of two DBR mirrors and a phase shifting dielectric metasurface layer. The metasurface is made of a uniform array of square cross section nano-posts.
  (e) Simulated transmission spectra of a set of filters as shown in (d) with different nano-post widths.}
  \label{2_phase}
\end{figure}

To design the FP filters, we first simulate and design transmissive dielectric metasurfaces using the rigorous coupled wave analysis (RCWA) technique \cite{Liu:2012gc}. We use transmissive dielectric metasurfaces that consist of amorphous silicon (\aSi{}, $n=3.40$) \mbox{nano-posts} on a square lattice (period: 600\,nm, height: 400\,nm) embedded in low-index SU-8 ($n=1.57$). The metasurface parameters are determined for achieving a large variation in the transmission phase by changing the width of the \mbox{nano-posts}, while the transmission is high enough within the wavelength range from 1450\,nm to 1700\,nm, as plotted in Fig.~\ref{2_phase}(a,b).
We use DBRs as the high reflectivity mirrors forming the FP resonator.
Each of the DBRs consists of 4 pairs of \aSi{} and SiO$_2$ ($n=1.47$) quarter-wavelength stacks. The simulated reflection spectrum of such a DBR is plotted in Fig.~\ref{2_phase}(c), and shows a stop-band in the range of $\Delta \lambda \sim 300\,\mbox{nm}$ around $\lambda=1550\,\mbox{nm}$ with reflectivities $R>0.99$.
When the cavity thickness is a half integer multiple of wavelength divided by the cavity refractive index, the FP resonance is formed inside the cavity and allows a single Lorentzian shaped peak in the transmission spectrum.
For this work, we chose the longitudinal mode number of 3 and found a single resonance within the DBR's stopband when the spacing between the DBR mirrors was filled with $\sim 1.2$-$\mu$m-thick SU-8 polymer.
Then, we incorporated the metasurface layers inside the SU-8 cavity layer to introduce the phase shift, and thus shift the resonance wavelengths of the FP resonators without changing the physical distances between the mirrors (Figure~\ref{2_phase}(d)).
We used the transfer matrix formalism to calculate the transmission spectra for a set of filters, using the complex transmission/reflection coefficients for the metasurface layers obtained via the RCWA simulations.


In Figure\,\ref{2_phase}(e), the simulated transmission spectra for a set of designed filters are plotted.
For this set, the widths of the \aSi{} \mbox{nano-posts} range from 120\,nm to 430\,nm.
By changing the widths of nano-posts array, the resonance wavelengths of the bandpass filters vary from 1450\,nm to 1700\,nm, spanning a 250\,nm bandwidth ($\Delta\lambda/\lambda_{\mathrm{c}}=16\%$), while the physical distance between the two mirrors in each filter is fixed. The planar form of these filters allows their fabrication using a single binary lithography step.
Each of the filters has a high transmission around the passband due to the low loss materials used in the designed nano-post metasurfaces.
The square cross section of the \mbox{nano-posts} and the square form of the lattice lead to the polarization insensitivity of the metasurface layer and the filters.


\section{Method and measurement results} 
\label{sec:method}

\subsection{Fabrication} 
\label{sub:fabrication}

\begin{figure}[p!]
  \centering
  \includegraphics{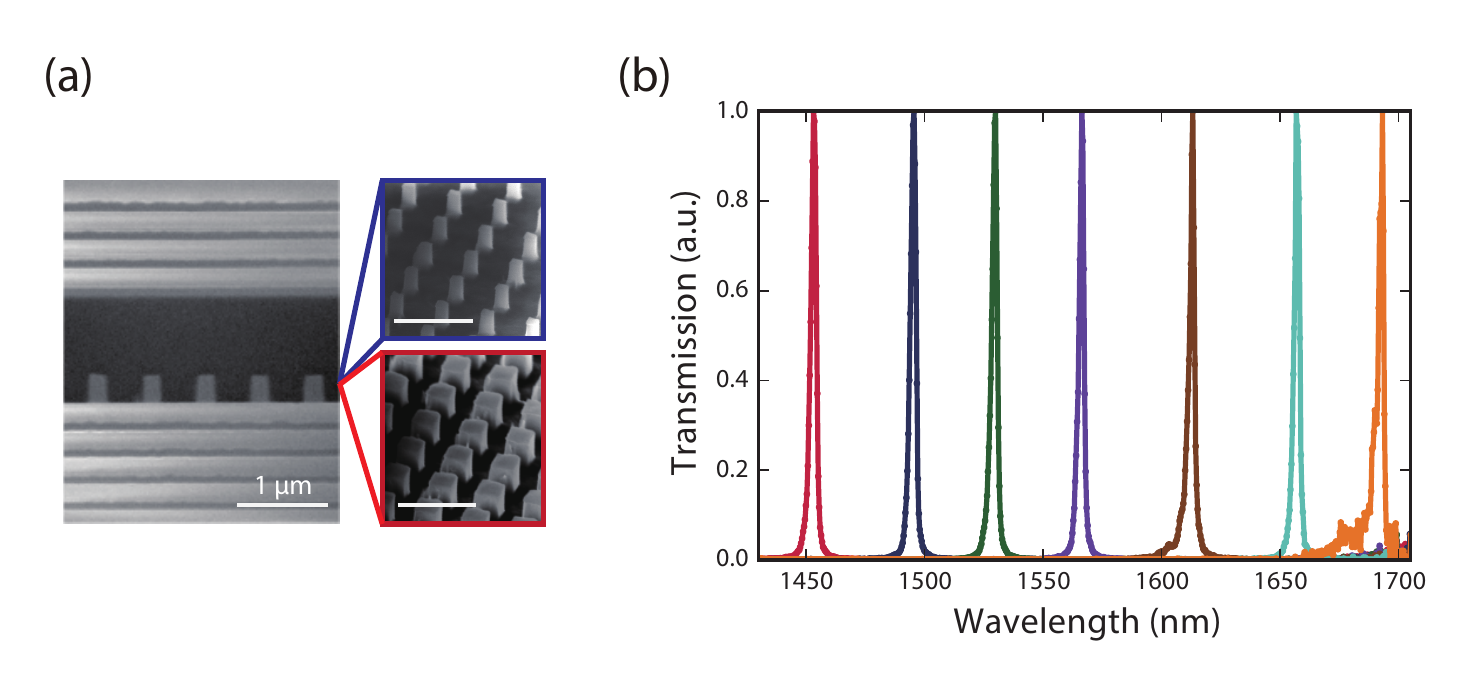}
  \caption{
  (a) A cross-sectional scanning electron microscope image of a fabricated filter, and bird's-eye views of two \aSi{} \mbox{nano-post} arrays with different widths. Scale bars are all $1\,\mu\mathrm{m}$.
  (b) Measured transmission spectra for the set of fabricated bandpass filters.
  The average measured absolute transmission for the filters is 16\% and their quality factors are around 700.
  }
  \label{fig_meas}
\end{figure}

As a proof of concept, the designed set of filters was fabricated on a single fused silica substrate.
First, the bottom DBR layers, a \mbox{258-nm-thick}  SiO$_2$ spacer layer and a \mbox{400-nm-thick} \aSi{} layer were deposited by plasma-enhanced chemical vapor deposition (PECVD).
The \mbox{nano-post} patterns were defined by electron beam lithography, first transferred into an aluminum oxide hard mask using a lift-off technique, and then to the \aSi{}  layer by dry etching \cite{Arbabi:2015iq}.
Then, \mbox{SU-8} polymer was spun and hard-baked, planarizing the entire area on the substrate \cite{Horie:2015bs}.
Finally, the top DBR layers were deposited by PECVD over the planar \mbox{SU-8} layer.
The cross-sectional scanning electron microscope image of the fabricated structure is shown in Fig.~\ref{fig_meas}(a). Two bird's-eye views of the \aSi{} \mbox{nano-posts} with two different widths before spinning the \mbox{SU-8} are also shown in Fig.~\ref{fig_meas}(a) as insets.
Each of the DBRs were composed of 4 alternating pairs of \aSi{} and SiO$_2$ layers (\aSi{} layers: 112\,nm, SiO$_2$ layers: 258\,nm).


\subsection{Characterization} 
\label{sub:characterization}

The fabricated bandpass filter array was characterized by measuring the transmission spectra of the filters.
A supercontinuum source was focused onto the fabricated filters using an objective lens, the transmitted light was collected using another objective lens, and its spectrum was measured using an optical spectrum analyzer.
The normalized transmission spectra were calculated by measuring the spectrum without the sample.
Fig.~\ref{2_phase}(e) shows the simulated transmission spectra computed for the fabricated designs, and Fig.~\ref{fig_meas}(b) shows the measured transmission spectra for the corresponding set of the filters with normalization.
The resonance wavelengths show good agreement between their simulated and measured values.
The measured quality factor was $\sim700$, and the measured absolute transmission power was 16\% in average.
The transmission measured from the area having no \aSi{} \mbox{nano-posts} shows similar peak transmission values and quality factors, indicating that the relatively low transmission and the measured linewidths of the filters are due to the loss from the fabricated DBRs.
The deposited DBR layers have significant surface roughness leading to scattering loss.
Optimization of the PECVD deposition conditions is expected to reduce the surface roughness of the deposited layers and improve the quality factors as well as the transmission power of the filters.



\section{Conclusion} 
\label{sec:conclusion}

We proposed and experimentally demonstrated a planar bandpass filter array based on vertical FP resonators. The filters have Lorentzian shape passband, and the center wavelength of each filter can be independently controlled by changing the in-plane dimensions of a low-loss dielectric metasurface layer inserted between two reflectors.
The planar geometry and the compatibility with binary lithography process, as well as the polarization insensitivity and large bandwidth of the proposed filter array make it ideal for implementation of low cost on-chip spectrometers with high resolving powers.


\section*{Acknowledgment}

This work was supported by Samsung Electronics and DARPA.
Y.~H. was supported by a Japan Student Services Organization (JASSO) fellowship.
S.~M.~K. was supported by the DOE Light-Material Interactions in Energy Conversion Energy Frontier Research Center funded by the US Department of Energy, Office of Science, Office of Basic Energy Sciences under Award no. DE-SC0001293.
The device nanofabrication was performed in the Kavli Nanoscience Institute at the California Institute of Technology.

\bibliography{DBRref}


\onecolumngrid

\end{document}